\documentclass[usenatbib]{mn2e}  
\usepackage{natbib}
\usepackage{epsfig}

\catcode`\@=11
\def\gta{\ifmmode{\,\mathrel{\mathpalette\@versim>\,}}
    \else{$\,\mathrel{\mathpalette\@versim>}\,$}\fi}
\def\lta{\ifmmode{\,\mathrel{\mathpalette\@versim<\,}}
    \else{$\,\mathrel{\mathpalette\@versim<}\,$}\fi}
\def\@versim#1#2{\lower 2.9truept \vbox{\baselineskip 0pt \lineskip
    0.5truept \ialign{$\m@th#1\hfil##\hfil$\crcr#2\crcr\sim\crcr}}}
\catcode`\@=12  
\renewcommand{\[}{\begin{equation}}
\renewcommand{\]}{\end{equation}}
\let\boldgrk=\gkvecten
\let\boldgrksc=\gkvecseven

\def\gkthing#1{{\mathchoice%
	{\hbox{{\boldgrk\char#1}}}
	{\hbox{{\boldgrk\char#1}}}
	{\hbox{{\boldgrksc\char#1}}}
	{\hbox{{\boldgrksc\char#1}}}}}

\def\vmu{\gkthing{22}}

{\newif\ifnotend
\notendtrue
\def\veclist{ABCDEFGHIJKLMNOPQRSTUVWXYZabcdefghijklmnopqrstuvwxyz.}
\def\top#1#2.{#1}
\def\tail#1#2.{#2.}
\loop\expandafter\xdef\csname v\expandafter\top\veclist\endcsname%
{{\noexpand\bf\expandafter\top\veclist}}
\edef\veclist{\expandafter\tail\veclist}
\if\veclist.\notendfalse\fi\ifnotend\repeat}
{\newif\ifnotend
\notendtrue
\def\veclist{ABCDEFGHIJKLMNOPQRSTUVWXYZ.}
\def\top#1#2.{#1}
\def\tail#1#2.{#2.}
\loop\expandafter\xdef\csname c\expandafter\top\veclist\endcsname%
{{\noexpand\cal\expandafter\top\veclist}}
\edef\veclist{\expandafter\tail\veclist}
\if\veclist.\notendfalse\fi\ifnotend\repeat}

\def\df{{\sc df}}

\def\Teff{T_{\rm eff}}

\def\vlos{v_{\rm los}}

\def\ex#1{\left\langle#1\right\rangle}

\def\mas{\,{\rm mas}}

\def\kms{\,{\rm km}\,{\rm s}^{-1}}

\def\yr{\,{\rm yr}}
\def\Gyr{\,{\rm Gyr}}
\def\K{\,{\rm K}} 
\def\pc{\,{\rm pc}}
\def\kpc{\,{\rm kpc}}

\def\e{{\rm e}}
\def\rd{{\rm d}}

\def\figref#1{Fig.~\ref{#1}}
\newcommand{\beq}{\begin{equation}}
\newcommand{\eeq}{\end{titleequation}}

\title[Galactic kinematics \& dynamics of RAVE stars]
{Galactic kinematics and dynamics from RAVE stars}

\author[J. Binney et al.]{J. Binney$^1$\thanks{E-mail:
binney@thphys.ox.ac.uk}, B. Burnett$^1$, G. Kordopatis$^2$, 
M. Steinmetz$^3$, G. Gilmore$^2$,\newauthor
O. Bienaym\'e$^4$, 
J. Bland-Hawthorn$^5$,
B. Famaey$^4$
E.K. Grebel$^6$,
A. Helmi$^7$, \newauthor
J. Navarro$^8$, 
Q. Parker$^{9,10}$,
W.A. Reid$^{9}$, 
G. Seabroke$^{11}$, 
A. Siebert$^4$, \newauthor
F. Watson$^{10}$,
M.E.K. Williams$^3$, 
R.F.G. Wyse$^{12}$
T. Zwitter$^{13}$
\\
$^1$ Rudolf Peierls Centre for Theoretical Physics, Keble Road, Oxford OX1
3NP, UK\\ 
$^2$ Institute of Astronomy, Madingley Road, Cambridge CB3 0HA, UK\\
$^3$ Leibniz-Institut für Astrophysik Potsdam (AIP), An der Sternwarte 16,
14482 Potsdam, Germany\\ 
$^4$ Observatoire Astronomique de Strasbourg, 11 rue de l'Universit\'e,
Strasbourg, France\\
$^5$ Sydney Institute for Astronomy, University of Sydney, School of Physics
A28, NSW 2006, Australia\\
$^{6}$ Astronomisches Rechen-Institut, Zentrum f\"ur Astronomie der
Universit\"at Heidelberg, M\"onchhofstr 12-14,\\ D-69120, Heidelberg,
Germany\\
$^{7}$ Kapteyn Astronomical Institut, University of Groningen, Landleven
12, 9747 AD, Groningen, The Netherlands\\
$^8$ Senior ClfAR Fellow, University of Victoria, BC Canada V8P 5C2 \\
$^{9}$ Macquarie University, Balaclava Road, NSW 2109, Australia\\
$^{10}$ Australian Astronomical Observatory, PO Box 915, North Ryde NSW
1670, Australia\\
$^{11}$ Mullard Space Science Laboratory, University College London,
Holmbury St Mary, Dorking, RH5 6NT, UK\\
$^{12}$Johns Hopkins University, Departement of Physics and Astronomy, 366
Bloomberg center, 3400 N. Charles St.,\\  Baltimore, MD 21218, USA\\
$^{13}$ University of Ljubljana, Faculty of Mathematics and Physics, Jadranska
19, 1000 Ljubljana, Slovenia and\\ Center of Excellence SPACE-SI, 
A\v{s}ker\v{c}eva cesta 12, 1000, Ljubljana, Slovenia\\
}

\begin{document}

\date{Draft, November 14 2013}

\pagerange{\pageref{firstpage}--\pageref{lastpage}} \pubyear{2012}

\maketitle

\label{firstpage}

\begin{abstract}

We analyse the kinematics of $\sim400\,000$ stars that lie within $\sim2\kpc$
of the Sun and have spectra measured in  the RAdial
Velocity Experiment (RAVE). We decompose the sample
into hot and cold dwarfs, red-clump and non-clump giants. The kinematics of
the clump giants are consistent with being identical with those of the giants
as a whole.  Without binning the data we fit Gaussian velocity ellipsoids to
the meridional-plane components of velocity of each star class and give
formulae from which the shape and orientation of the velocity ellipsoid can
be determined at any location. The data are consistent with the giants and
the cool dwarfs sharing the same velocity ellipsoids, which have vertical
velocity dispersion rising from $21\kms$ in the plane to $\sim55\kms$ at
$|z|=2\kpc$ and radial velocity dispersion rising from $37\kms$ to $82\kms$
in the same interval. At $(R,z)$ the longest axis of one of these velocity
ellipsoids is inclined to the Galactic plane by an angle
$\sim0.8\arctan(z/R)$.  We use a novel formula to obtain precise fits to the
highly non-Gaussian distributions of $v_\phi$ components in eight bins in the
$(R,z)$ plane.  

We compare the observed velocity distributions with the predictions of a
published dynamical model fitted to the velocities of stars that lie within
$\sim150\pc$ of the Sun and star counts towards the Galactic pole. The
predictions for the $v_z$ distributions are exceptionally successful. The
model's predictions for $v_\phi$ are successful except for the hot dwarfs,
and its predictions for $v_r$ fail significantly only for giants that lie far
from the plane.  If distances to the model's stars are over-estimated by 20
per cent, the predicted distributions of $v_r$ and $v_z$ components become
skew, and far from the plane broader. The broadening significantly improves
the fits to the data.

The ability of the dynamical model to give such a good account of a large
body of data to which it was not fitted inspires confidence in the
fundamental correctness of the assumed, disc-dominated, gravitational
potential.

\end{abstract}


\section{Introduction}

A major strand of contemporary astronomy is the quest for an understanding of
how galaxies formed and evolved within the context of the concordance
cosmological model, in which the cosmic energy density is dominated by vacuum
energy and the matter density is dominated by some initially cold matter that
does not interact electromagnetically. This quest is being pursued on
three fronts: observations of objects seen at high redshifts and early times,
simulations of clustering matter and star formation, and by detailed
observation of the interplay between the chemistry and dynamics of stars in
our own Galaxy. 

As a contribution to this last ``Galactic archaeology'' strand of the quest
for cosmic understanding, the RAdial Velocity Experiment
\citep{SteinmetzRAVE} has since 2003 gathered spectra at resolution
$\sim7500$ around the CaII near-IR triplet of $\sim400\,000$ stars. The
catalogued stars are roughly half giants and half dwarfs, and mostly lie
within $2.5\kpc$ of the Sun \citep{Burnettetal11,Binneyetal13}. The RAVE survey is
complementary to the Sloan Digital Sky Survey \citep[SDSS;][]{York} and the
latter's continuations \citep{Yanny,Eisenstein} in that it observes stars at
least as bright as $I=9-13$, whereas the SDSS observes stars fainter than
$g=14$. On account of the faint magnitudes of the SDSS stars, they are
overwhelmingly at distances greater than $0.5\kpc$ so the Galaxy's thin disc,
which has a scale height $\sim0.3\kpc$ and is by far the dominant stellar
component of the Galaxy, contributes a small proportion of the stars in the
SDSS data releases.  The thin and thick discs, by contrast, completely
dominate the RAVE catalogue.

Recently \cite{Binneyetal13} derived distances to $\sim400\,000$ stars from
2MASS photometry and the stellar parameters produced by the VDR4
spectral-analysis pipeline described by \cite{Kordopatis12}. We use these
distances to discuss the kinematics of the Galaxy in the extended solar
neighbourhood, that is, in the region within $\sim2\kpc$ of the Sun. Since
the selection criteria of the RAVE survey are entirely photometric, we can
determine the distribution of the velocities of survey stars within the
surveyed region without determining the survey's complete selection function,
which is difficult (see Piffl \& Steinmetz in preparation, Sharma et al in
preparation).

We characterise the kinematics in several distinct ways. In Section
\ref{sec:nobins} we obtain analytic fits to the variation within the $(R,z)$
plane of the velocity ellipsoid by a technique that avoids binning stars
\citep{Burnett10}. In Section \ref{sec:bins} we bin stars to obtain
histograms of the distribution of three orthogonal components of velocity. We
use a novel formalism to obtain analytic fits to the distributions of the
azimuthal component of velocity.  We examine the first and second moments of
the distributions of the velocity components parallel to the principal axes
of the local velocity ellipsoid. The second moments are
consistent with our previously derived values, but some first moments are
non-zero: values $\sim1.5\kms$ are common and values as large as $5\kms$
occur.

In Section \ref{sec:fjmodels} we compare our results with the predictions of
a dynamical model Galaxy that is based on Jeans' theorem. Although this
model, which was described by Binney (2012; hereafter B12), was not fitted to
any RAVE data, we find that its predictions for the distributions of vertical
components are extremely successful, while those for the radial components
are successful at $|z|<0.5\kpc$ but become less successful further from the
plane, where they produce velocity distributions that are too narrow and
sharply peaked. In Section \ref{sec:errors} we investigate the impact of
systematically over-estimating distances to stars. When distances to the
model's stars are over-estimated by 20\%, the predicted distributions of
$v_r$ and $v_z$ acquire asymmetries that are similar to those sometimes seen
in the data. Systematic over-estimation of distances brings the model into
better agreement with data far from the plane by broadening its $v_r$
distributions.

\section{Input parameters and data}

Throughout the paper we adopt $R_0=8\kpc$ as the distance of the Sun from the
Galactic centre, $\Theta_0=220\kms$ for the local circular speed and from
\cite{SchoenrichBD} $(U_0,V_0,W_0)=(11.1,12.24,7.25)\kms$ as the velocity of
the Sun with respect to the Local Standard of Rest. While our values of $R_0$
and $\Theta_0$ may be smaller than they should be \citep[e.g.][]{McMillan11},
we adopt these values in order to be consistent with the assumptions inherent
in the B12 model.

Proper motions for RAVE stars can be drawn from several catalogues.
\cite{Williams13} compares results obtained with different proper-motion
catalogues, and on the basis of this discussion we originally decided to work
with the PPMX proper motions \citep{Roeseretal} because these are available
for all our stars and they tend to minimise anomalous streaming motions.
However, when stars are binned spatially and one computes the dispersions in
each bin of the raw velocities $4.73\vmu (s/\hbox{kpc})+\vv_{\rm los}$ from
the PPMX proper motions, the resulting dispersions for bins at distances
$\gta0.5\kpc$ are often smaller than the contributions to these from
proper-motion errors alone. It follows that either our distances are much too
large, or the quoted proper-motion errors are seriously over-estimating the
true random errors. The problem can be ameliorated by cutting the sample to
exclude stars with large proper-motion errors, but there are still signs that
the velocity dispersions in distant bins are coming out too small on account
of an excessive allowance for the errors in the proper motions of stars that
have small proper motions. The errors in the UCAC4 catalogue \citep{UCAC4}
are $\sim60$ percent of those in the PPMX catalogue and the problem just
described does not arise with these proper motions, so we have used them. We
do, however, exclude stars with an error in $\mu_b$ greater than
$8\mas\yr^{-1}$.

In addition to this cut on proper-motion error, the sample is restricted to
stars for which \cite{Binneyetal13} determined a probability density
function (pdf) in distance modulus. To
belong to this group a star has to have a spectrum that passed the
\cite{Kordopatis12} pipeline with S/N ratio of 10 or more.

\section{Fitting meridional components without binning the data}\label{sec:nobins}

At each point in the Galaxy a stellar population that is in statistical
equilibrium in an axisymmetric gravitational potential $\Phi(R,z)$ should
define a velocity ellipsoid. Two of the principal axes of this ellipsoid
should lie within the $(R,z)$ plane, with the third axis in the azimuthal
direction $\ve_\phi$. Near the plane the ellipsoid's longest axis is expected
to point roughly radially and the shortest axis vertically. Let $\ve_1$ be
the unit vector along the longest axis, and $\ve_3$ be the complementary unit
vector, and let $\theta(R,z)$ denote the angle between $\ve_1$ and the
Galactic plane.

The lengths of the principal semi-axes of the velocity ellipsoid are of course the
principal velocity dispersions
\begin{eqnarray}
\sigma_1(R,z)&=&\ex{(\vv\cdot\ve_1)^2}^{1/2}\nonumber\\
\sigma_\phi(R,z)&=&\left(\ex{(\vv\cdot\ve_\phi)^2}-\ex{\vv\cdot\ve_\phi}^2\right)^{1/2}\\
\sigma_3(R,z)&=&\ex{(\vv\cdot\ve_3)^2}^{1/2}.\nonumber
\end{eqnarray}
In the following we shall use the notation
 \[\label{eq:defsV1}
V_1\equiv\vv\cdot\ve_1\hbox{ and } V_3\equiv\vv\cdot\ve_3.
\]
 We estimate the functional
forms of $\sigma_1$ and $\sigma_3$ as follows. 

\begin{figure}
\centerline{\epsfig{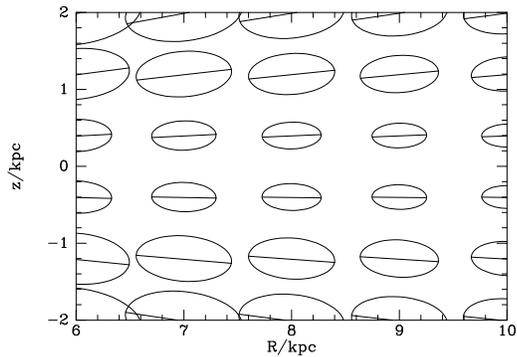}}
\caption{Representation of the velocity ellipsoids of giant
stars; the lengths of the principal axes of each ellipse are proportional to
the corresponding velocity dispersion at the centre of the ellipse.}\label{fig:ellipsoids}
\end{figure}

 We let $\theta(R,z)$ be determined by a single parameter $a_0$ through
 \[\label{eq:defa0}
\theta=a_0\arctan\left(z/ R\right).
\]
 We use four further parameters $a_i$ to constrain the behaviour of $\sigma_1$, and
similarly for $\sigma_3$, by writing
 \begin{eqnarray}\label{eq:defsmod}
\sigma_1(R,z)&=&\sigma_0a_1\exp[-a_2(R/R_0-1)][1+(a_3z/R)^2]^{a_4}\nonumber\\
\sigma_3(R,z)&=&\sigma_0a_5\exp[-a_6(R/R_0-1)][1+(a_7z/R)^2]^{a_8},
\end{eqnarray}
 where $\sigma_0\equiv 30\kms$ ensures that all the $a_i$ are dimensionless
and of order unity.  These forms are the fruit of a combination of physical
intuition and some experimentation. In particular, by symmetry we require
even functions of $z$ that have vanishing vertical gradients in the plane,
and experimentation shows that power series in $z^2$ do not work well.
Second, it has been conventional to assume exponential dependence of velocity
dispersion on $R$ since the scale heights of discs were found to be roughly
constant \citep{vanderK}. Moreover, the data cover a significant range in $R$
only at large $|z|$, so we are not in a position to consider elaborate
dependence on $R$. The parameters $a_1$ and $a_5$ set the overall velocity
scale of $\sigma_1$ and $\sigma_3$, respectively, while $a_2$ and $a_6$
determine how fast these dispersions decrease with increasing radius. The
parameter pairs $(a_3,a_4)$ and $(a_7,a_8)$ determine how the dispersions
vary with distance from the plane.
 
\begin{table*}
\begin{center}
\caption{Test of the fitting procedure. The bottom row gives the parameters
used to choose the velocities, while top row gives the values of the
parameters in equation (\ref{eq:defsmod}) from which {\sc frprmn} started.
The second row shows the values of the parameters on which it converged given
data at the locations of the 40\,175 clump giants. The third, fourth and
fifth rows
give the parameters values similarly obtained using data at the locations of
181\,725 non-clump giants, 55\,398 hot dwarfs and 95\,470 cool dwarfs,
respectively.}\label{tab:test}
\begin{tabular}{lccccccccc}
&$a_0$&$a_1$&$a_2$&$a_3$&$a_4$&$a_5$&$a_6$&$a_7$&$a_8$\\
\hline
start&1&.5&0.1&2&1&1&0.2&5&1\\
Clump giants& 0.506 & 1.011 & 0.414 & 5.355 & 0.549 & 0.493 & 0.307 &11.425 & 0.433\\
Non-clump giants& 0.491 & 0.998 & 0.482 & 6.519 & 0.462 & 0.499 & 0.347 & 9.768 & 0.511\\
Hot dwarfs& 0.459 & 0.994 & 0.611 & 3.329 & 2.194 & 0.499 & 0.448 & 5.241 & 1.598\\
Cool dwarf& 0.587 & 1.003 & 0.541 & 2.905 & 1.841 & 0.499 & 0.210 & 5.505 & 1.500\\
truth&0.5&1&0.4&6&0.5&0.5&0.4&10&0.5\\
\hline
\end{tabular}
\end{center}
\end{table*}

\begin{figure}
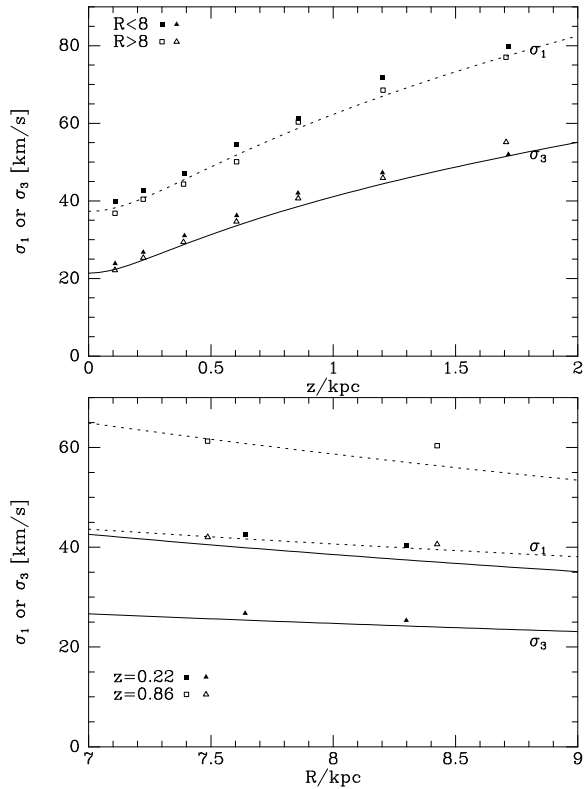

\centerline{\epsfig{file=gsr.ps,width=.9\hsize}}
\centerline{\epsfig{file=gsz.ps,width=.9\hsize}}
 \caption{The curves show the spatial variation of the values of $\sigma_3$
and $\sigma_1$ at fixed $R$ (top) or $z$ (below) that are extracted from the
raw data for non-clump giant stars by a maximum-likelihood technique that takes into
account random measuring errors. The black dots show the result of correcting
the dispersions of binned data for measurement errors by simple quadrature
subtraction. In the upper panel the upper point of each pair refers to a bin
that lies inside $R_0$ and the lower point refers to a bin at $R>R_0$. In the
lower panel results are shown for $z=0.22$ and
$0.86\kpc$.}\label{fig:gSigzSigR}
\end{figure}

From equations (\ref{eq:defsmod}) it is straightforward to calculate the
derivatives with respect to the nine parameters $a_i$ of the components
$V_1,V_3$ of a star's velocity and of the dispersions $\sigma_i$, so
we use a conjugate-gradient method to extremise the log-likelihood
 \[\label{eq:logL}
\sum_{\rm stars}\sum_{i=1,3}\ln[\sigma_i^2+e^2(V_i)]+{V_i^2\over
\sigma_i^2+e^2(V_i)},
\]
 associated with a correctly normalised biaxial Gaussian pdf
in $(V_1,V_2)$ space. Here $e(V_i)$ is the formal error in $V_i$ for a given
star. This is computed from the quoted errors on the proper motions and the
line-of-sight velocity assuming the distance to be inverse of the expectation
of the parallax given by \cite{Binneyetal13}, who found this to be the most
reliable distance estimator. With the present method it is exceedingly hard
to allow for distance errors, and we do not do this.

The code for extracting the values of the $a_i$ from a catalogue of stellar
phase-space coordinates was tested as follows. The velocity of each RAVE star
was replaced by a velocity chosen at random from a triaxial Gaussian velocity
distribution with variances $\sigma_i^2(R,z)+e^2(V_i)$, where the $\sigma_i$
were derived from plausible values of the $a_i$ and the errors $e(V_i)$ are
the actual errors on that star's velocity components.  Then the routine {\sc
frprmn} of \cite{Pressetal} was used to maximise the function (\ref{eq:logL})
starting from another set of values of the $a_i$.  The
conventional $\chi^2$ is
 \[
\chi^2=\sum_{{\rm stars}}\sum_{i=1,3}
{V_i^2\over\sigma_i^2+e^2(V_i)}.
\]
 In all tests the chosen model yielded a value of $\chi^2$ per degree of
freedom that differed from unity by less than $3\times10^{-4}$.

We have analysed separately four classes of stars: clump giants ($0.55\le
J-K\le 0.8$ and $1.7\le\log g<2.4$), non-clump giants ($\log g<3.5$),
hot ($\Teff>6000\K$) dwarfs and cool dwarfs.  

The first row of Table~\ref{tab:test} shows the parameters from which fitting
started, while the bottom row gives the values of the parameters that were
used to assign velocities to the stars. The second row shows the parameter
values upon which {\sc frprmn} converged with data at the locations of
40\,175 red-clump stars in the RAVE sample. The third row gives the results
obtained using the sample's 181\,726 non-clump giants. The
fourth and fifth rows give, respectively, results obtained using the 55\,398
hot dwarfs and 95\,469 cool dwarfs. 

Naturally the precision with which the parameters can be recovered from the
data increases with the size and spatial coverage of the sample. Hence the
cold dwarfs deliver the least, and the giants the most, accurate results. The
parameters that are most accurately recovered are $a_1$ and $a_5$, which
control the magnitudes of dispersions, and $a_0$, which controls the tilt of
the velocity ellipsoid.  The parameters $a_3$ and $a_4$, which control the
vertical variation of the radial dispersion, and $a_7$ and $a_8$, which
control the vertical variation of the vertical dispersion, are recovered
quite well from the giants but rather poorly from the dwarfs. However, even
the dwarfs yield quite accurate values for the products $a_3^2\,a_4$ and
$a_7^2\,a_8$ that occur in the first non-trivial term in the Maclaurin series
of the final brackets of equations (\ref{eq:defsmod}). The parameters $a_2$
and $a_6$, which control radial gradients are recovered only moderately well
by all star classes.

\begin{table*}
\begin{center}
\caption{Velocity ellipsoids from measured velocities. When the values given
here are inserted into equations (\ref{eq:defa0}) and (\ref{eq:defsmod}) one
obtains expressions for the semi-axis lengths and orientation of the velocity
ellipsoids at a general point $(R,z)$.  From top to bottom the rows give
results for clump giants, non-clump giants, and hot and cool
dwarfs.}\label{tab:SigzSigR}
\begin{tabular}{lccccccccc}
&$a_0$&$a_1$&$a_2$&$a_3$&$a_4$&$a_5$&$a_6$&$a_7$&$a_8$\\
\hline
Clump giants& 0.872 & 1.183 & 0.394 &24.835 & 0.212 & 0.682 & 0.554 &29.572 & 0.211\\
Non-clump giants& 0.815 & 1.243 & 0.398 &25.283 & 0.214 & 0.713 & 0.362 &34.815 & 0.218\\
Hot dwarfs& 0.213 & 0.976 & 0.719 & 7.891 & 1.282 & 0.468 &-0.209 &26.992 & 0.380\\
Cool dwarfs& 0.815 & 1.153 & 1.142 &47.112 & 0.169 & 0.711 & 1.572 & 9.852 & 1.200\\
\hline
\end{tabular}
\end{center}
\end{table*}

When fitting the measured velocities of RAVE stars, the difference between unity
and $\chi^2$ per degree of freedom for the chosen model ranged from
$3.5\times10^{-3}$ for cold dwarfs to $1.7\times10^{-2}$ for non-clump
giants.  Table~\ref{tab:SigzSigR} shows the parameters of the chosen models.
Both classes of giants and the cool dwarfs yield similar values $a_0\simeq0.8$
of the parameter that controls the orientation of the velocity ellipsoid.
Since this value lies close to unity, the long axis of the velocity ellipsoid
points almost to the Galactic centre (\figref{fig:ellipsoids}) consistent with the findings of
\cite{Siebert08}. The hot dwarfs yield a much smaller value, $a_0\simeq0.2$,
so the long axis of their velocity ellipsoid does not tip strongly as one
moves up. 

The velocity dispersions in the plane are $\sigma_R=30a_1\kms$ and
$\sigma_z=30a_5\kms$. The smallest dispersions,
$(\sigma_R,\sigma_z)=(29.3,14.0)$ are for the hot dwarfs and the largest,
$(37.3,21.4)$ are for the giants.  For the giants and cool dwarfs we have
$\sigma_z/\sigma_R=a_5/a_1\simeq0.6$, while for the hot dwarfs we have
$\sigma_z/\sigma_R\simeq0.48$, significantly smaller. 

The scale lengths on which the dispersions vary are $R_\sigma=R_0/a_2$ for
$\sigma_r$ and $R_\sigma=R_0/a_6$ for $\sigma_z$. For the giants these are
$\sim2.5 R_0$, which is surprisingly large: one anticipates
$R_\sigma\lta3R_\rd\simeq R_0$. The cool dwarfs, by contrast yield
$R_\sigma<R_0$. For $\sigma_r$ the hot dwarfs yield $R_\sigma\simeq1.4R_0$,
but for $\sigma_z$ they yield a negative value of $R_\sigma$, implying that
$\sigma_z$ {\it increases\/} with radius.  Given that the survey volume is a
cone that excludes the plane, not only is it hard to disentangle radial and
vertical gradients, but stars such as hot dwarfs that are strongly
concentrated to the plane do not probe a large volume and consequently are
not suited to measuring gradients. Moreover, the longest axis of the velocity
ellipsoids of populations of young stars are known not to lie within the
$(R,z)$ plane -- the ``vertex deviation'' \citep[e.g.][]{DehnenB98}. This
phenomenon is evidence that these populations are  not in dynamical
equilibrium as our methodology assumes, either because they are too young, or
because they are strongly disturbed by spiral structure.

\begin{figure}
\centerline{\epsfig{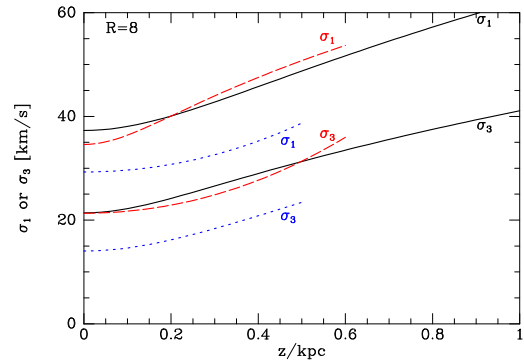}}
\caption{The dependence on $|z|$ of the velocity dispersions at $R_0$. Full black
curves are for non-clump giants, red dashed curves are for cool dwarfs and dotted blue
curves are for hot dwarfs.}\label{fig:allsig}
\end{figure}

The upper panel of \figref{fig:gSigzSigR} shows the dependencies on $z$ at
$R=8\kpc$ of $\sigma_1$ (dashed line) and $\sigma_3$ (full line) that are
implied by Table~\ref{tab:SigzSigR} for non-clump giants. The squares and
triangles show velocity dispersions estimated by binning the data as
described in Section \ref{sec:bins} below. The lower panel shows the
corresponding radial dependencies at $z=0.22$ and $z=0.86\kpc$.

In \figref{fig:allsig} the full black curves show the runs with $z$ at
$R=R_0$ of $\sigma_1$ and $\sigma_3$ for non-clump giants, while the dashed
red curves show the same quantities for the cool dwarfs. From these plots we
infer that the dispersions of the cool dwarfs are probably consistent with
those for non-clump giants except very near the plane where $\sigma_1$ may be
lower for the dwarfs. The blue dotted curves show the distinctly lower
velocity dispersions of the hot dwarfs: lower dispersions are to be expected
of such relatively young stars since they have experienced less stochastic
acceleration than older stars.

\section{Using binned data}\label{sec:bins}

\subsection{Azimuthal velocities}

In a disc galaxy, the distribution of $v_\phi$ components is inherently skew
and the skewness of the distribution contains essential information about the
system's history and dynamics. Consequently, it is not appropriate to use the
machinery described in the last section to fit observed $v_\phi$
distributions.

\begin{figure}
\centerline{\epsfig{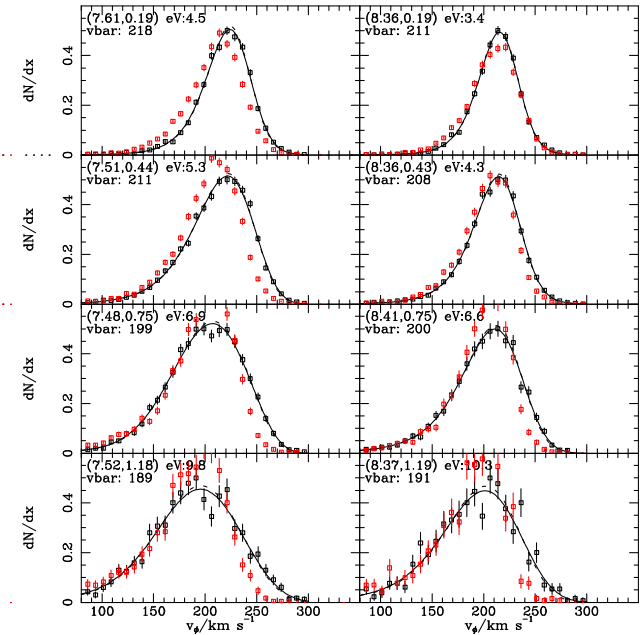}}
 \caption{The distributions of $v_\phi$ for red-clump giants (black data
points) and fits to them -- in each panel the dashed curve shows the
kinematic model specified by equations (\ref{eq:Pphi}) and (\ref{eq:Svphi}),
while the full curve shows the model convolved with the mean errors in
$v_\phi$. The red points show the predictions of the B12 dynamical model. The
mean coordinates of the stars in each bin are given at top left, followed by
the rms velocity error (eV) and the sample mean of $v_\phi$ (vbar). In this and all
subsequent histograms, the horizontal bars span the width of the bins and the
vertical bars indicate Poisson errors.}\label{fig:rcVphiFits}
\end{figure}

\begin{figure}
\centerline{\epsfig{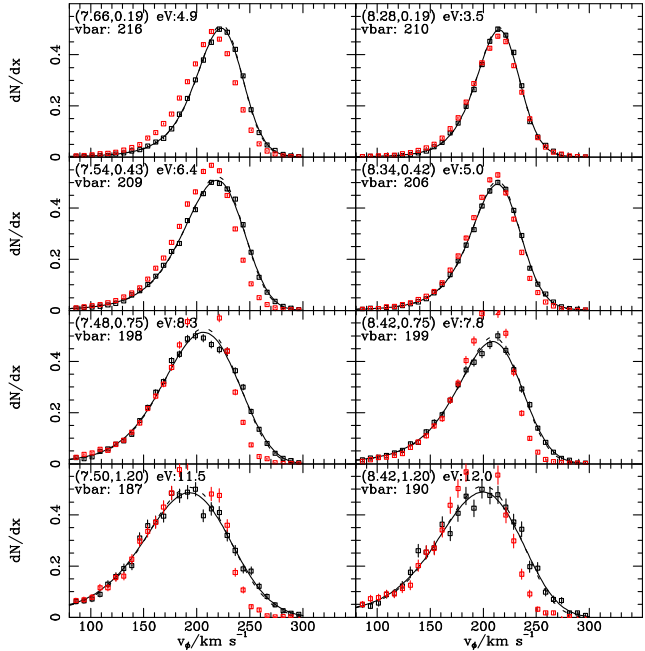}}
\caption{As \figref{fig:rcVphiFits} but for non-clump giant stars.
}\label{fig:gVphiFits}
\end{figure}

The $v_\phi$ distributions of the dynamical models described by B12, which
will be discussed in Section \ref{sec:fjmodels} below, can be fitted
extremely well by the following analytic distribution
 \[\label{eq:Pphi}
P(v_\phi)=\hbox{constant}\times\e^{-(v_\phi-b_0)^2/2\sigma_\phi^2},
\]
 where $\sigma_\phi$ is a cubic in $v_\phi$:
\[\label{eq:Svphi}
\sigma_\phi(v_\phi)=b_1+b_2v_{\phi100}+b_3v_{\phi100}^2+b_4v_{\phi100}^3,
\]
 with $v_{\phi100}\equiv v_\phi/100\kms$. The general idea here is that
$b_0$ defines a characteristic streaming velocity, while $b_1$ is a basic
azimuthal velocity dispersion. The parameters $b_2$ to $b_4$ cause the velocity
dispersion $\sigma_\phi$ to increase/decrease as $v_\phi$ moves below/above
the circular speed, thus making the $v_\phi$ distribution skew.

In principle functional forms could be adopted for the dependence on $(R,z)$
of the parameters $b_i$ appearing in equations (\ref{eq:Pphi}) and
(\ref{eq:Svphi}), and then, in strict analogy to the work of the previous
section, the values of the parameters appearing in these functional forms
could be determined by maximising the likelihood of the data given the
distribution (\ref{eq:Pphi}).  Unfortunately, for this scheme to be viable we
require an expression for the value of the normalising constant as a function
of the parameters, and no such formula is available. Therefore we have
determined the $b_i$ by binning the data and doing a least-squares fit of
equation (\ref{eq:Pphi}) convolved with the observational errors to the
histogram of the binned data.

\begin{table}
\begin{center}
\caption{Values of the mean streaming velocity and the parameters defined by equations (\ref{eq:Pphi}) and
(\ref{eq:Svphi}) required to fit the $v_\phi$ distributions of RAVE stars.
The upper block refers to red clump stars and the lower one  to non-clump
giants.}\label{tab:gVphi} 
\begin{tabular}{lcccccc}
$(R,|z|)$&$\overline{v}_\phi$&$b_0$&$b_1$&$b_2$&$b_3$&$b_4$\\
\hline
(7.61, 0.19)& 217.9& 224.2&  51.0&  -5.79&  -9.78&  2.81\\
(8.36, 0.19)& 211.4& 215.5&  45.1&  -2.48& -12.34&  3.60\\
(7.51, 0.44)& 210.8& 222.0&  58.6& -14.91&   0.07&  0.20\\
(8.36, 0.43)& 207.9& 214.7&  49.7&  -2.70& -12.22&  3.43\\
(7.48, 0.75)& 199.0& 207.3&  71.2& -50.09&  27.76& -5.55\\
(8.41, 0.75)& 200.1& 211.4&  62.4& -18.73&  -0.20&  0.63\\
(7.52, 1.18)& 189.3& 195.8&  71.2& -39.27&  18.50& -3.27\\
(8.37, 1.19)& 191.2& 201.9&  70.1& -30.49&   9.61& -1.44\\
\hline
(7.66, 0.19)& 215.6& 223.3&  53.6&  -4.15& -12.21&  3.36\\
(8.28, 0.19)& 209.8& 215.1&  52.8& -11.90&  -7.53&  2.74\\
(7.54, 0.43)& 208.7& 219.2&  63.8& -21.69&   1.61&  0.31\\
(8.34, 0.42)& 206.4& 213.5&  57.0& -12.40&  -7.85&  2.83\\
(7.48, 0.75)& 198.2& 206.7&  72.0& -41.92&  17.72& -2.96\\
(8.42, 0.75)& 198.7& 209.3&  66.1& -23.36&   1.59&  0.53\\
(7.50, 1.20)& 186.6& 193.3&  76.4& -42.35&  16.29& -2.28\\
(8.42, 1.20)& 190.2& 200.3&  78.0& -44.79&  18.81& -3.20\\
\hline
\end{tabular}
\end{center}
\end{table}

\begin{table}
\begin{center}
\caption{The same as Table~\ref{tab:gVphi} but for hot (upper block) and cool
(lower block) dwarfs.}\label{tab:dVphi}
\begin{tabular}{lcccccc}
$(R,|z|)$&$\overline{v}_\phi$&$b_0$&$b_1$&$b_2$&$b_3$&$b_4$\\
\hline
(7.85, 0.10)& 220.1& 224.9&  69.5& -44.33&  10.68& -0.73\\
(8.11, 0.11)& 216.5& 220.1&  29.3&  20.80& -24.86&  5.67\\
(7.80, 0.22)& 220.7& 224.4&  29.8&  20.98& -24.10&  5.36\\
(8.13, 0.22)& 217.5& 221.3&  29.6&  21.56& -25.23&  5.69\\
(7.78, 0.36)& 219.5& 225.0&  46.9&  -0.85& -13.53&  3.59\\
(8.15, 0.36)& 215.8& 219.2&  79.2& -56.54&  14.43& -0.71\\
(7.79, 0.50)& 217.6& 223.2&  46.8&  -3.40& -10.04&  2.75\\
(8.15, 0.50)& 214.3& 218.7&  69.6& -37.94&   5.23&  0.74\\
\hline
(7.90, 0.09)& 215.8& 222.2&  -9.6&  98.37& -66.58& 12.49\\
(8.06, 0.08)& 213.7& 219.9& -18.9& 111.09& -72.53& 13.42\\
(7.84, 0.21)& 211.1& 219.7&  18.8&  52.04& -41.09&  8.08\\
(8.10, 0.21)& 211.1& 217.6&  -4.6&  87.28& -59.59& 11.26\\
(7.81, 0.36)& 211.5& 219.9&  19.7&  58.80& -47.62&  9.56\\
(8.12, 0.35)& 207.7& 215.2&  57.3& -12.98&  -8.21&  2.90\\
(7.73, 0.50)& 203.6& 216.1&  22.4&  52.28& -39.99&  7.54\\
(8.16, 0.51)& 210.9& 218.4&   8.8&  87.40& -67.29& 13.41\\
\hline
\end{tabular}
\end{center}
\end{table}

\begin{figure}
\centerline{\psfig{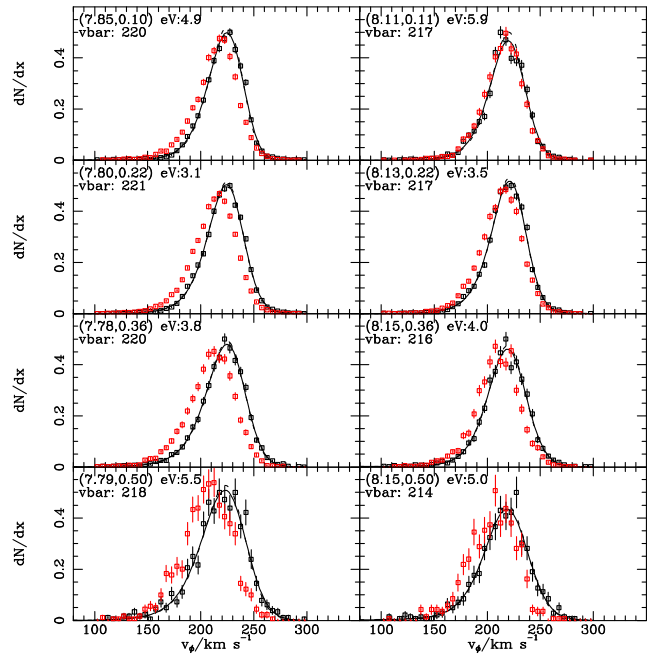}}
\caption{As \figref{fig:rcVphiFits} but for hot dwarfs.
}\label{fig:hdVphiFits}
\end{figure}

\begin{figure}
\centerline{\epsfig{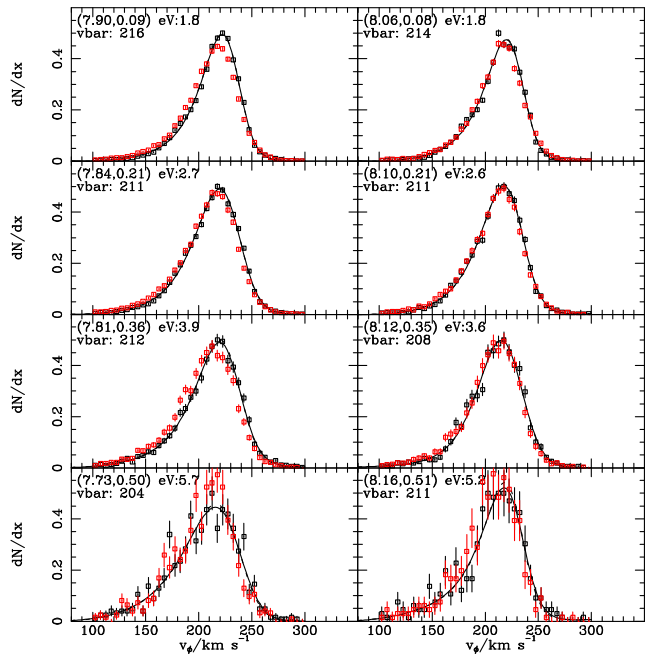}}
\caption{As \figref{fig:rcVphiFits} but for cool dwarfs.
}\label{fig:cdVphiFits}
\end{figure}

\begin{figure}
\centerline{\epsfig{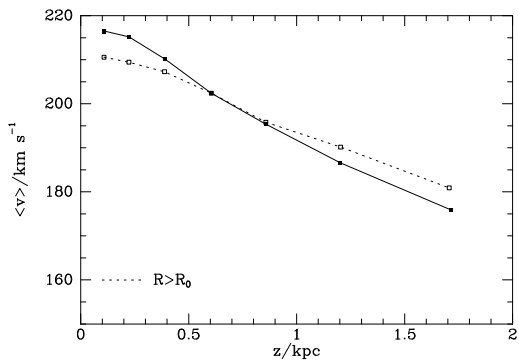}}
\caption{The mean rotation velocity of the giants as a function of distance
from the plane. The full curve is for bins at $R<R_0$. The data points are
the means of model distributions like those plotted as dotted curves in
Fig.~\ref{fig:gVphiFits}. The statistical errors on these points are very
small.}\label{fig:gvbar}
\end{figure}

\begin{figure}
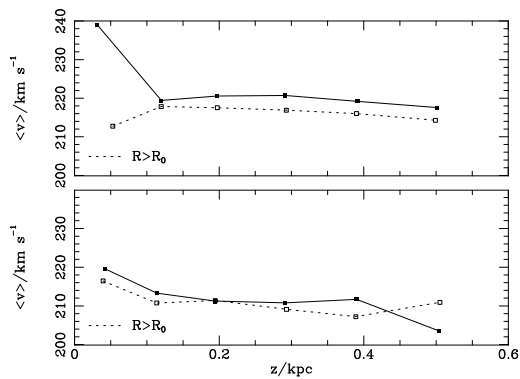

\hbox to \hsize{\qquad\epsfig{file=hd_vphibar.ps,width=.785\hsize}\hfil}
\hbox to \hsize{\qquad\epsfig{file=cd_vphibar.ps,width=.8\hsize}\hfil}
\caption{As \figref{fig:gvbar} but for the dwarfs: hot (top) and cool (below).}\label{fig:dvbar}
\end{figure}

The stars were divided into 8 spatial bins according to whether $R<R_0$ or
$R>R_0$ and $|z|$ lay in intervals bounded by $(0,0.3,0.6,1,1.5)\kpc$ for
giants or $(0,0.15,0.3,0.45,0.6)\kpc$ for dwarfs. Table~\ref{tab:gVphi} gives
the parameters that fit the $v_\phi$ distributions of the clump stars (upper
block) and non-clump giants (lower block). Table~\ref{tab:dVphi} gives values
of the parameters for the hot (upper block) and cool dwarfs.  The black
points in Figs.~\ref{fig:rcVphiFits} to \ref{fig:cdVphiFits} show the
observational histograms. At the top left of each panel we give the mean
values of $(R,|z|)$ and $e(v_\phi)$ for stars in the bin, where the latter is
the r.m.s. error for the stars in the given bin.  Also given at the top of
each panel is the mean velocity, $\ex{v_\phi}$, which of course is sensitive
to our adopted values $\Theta_0=220\kms$ and
$v_{\phi\odot}=\Theta_0+12.24\kms$.  The values of $\ex{v_\phi}$ are also
given in Tables~\ref{tab:gVphi} and \ref{tab:dVphi}, where we see that on
account of the skewness of the $v_\phi$ distributions, $\ex{v_\phi}$ is
systematically smaller than the fit parameter $b_0$, which would be the mean
velocity if $\sigma_\phi$ were not a function of $v_\phi$.

In Figs.~\ref{fig:rcVphiFits} to \ref{fig:cdVphiFits} bins with $R<R_0$ are
shown in the left column, bins with $R>R_0$ are shown in the right column,
and $|z|$ increases downwards.  The dotted curves show the functions defined
by the $b_i$ in Tables~\ref{tab:gVphi} and \ref{tab:dVphi} while the full
curves show the results of convolving these curves with the Gaussian of
dispersion $e(v_\phi)$. The dotted curves are mostly obscured by the full
curves because observational errors do not have a big impact on these data.
All histograms are fitted to great precision by the full curves.

Figs~~\ref{fig:gvbar} and \ref{fig:dvbar} show, respectively, the mean
rotation velocity of the giants and dwarfs as functions of distance from the
plane. The data points were obtained by fitting the analytic model convolved
with the measurement errors to
histograms of $v_\phi$ components with the stars placed in seven bins at each
of $R<R_0$ and $R>R_0$, and then calculating for each bin the mean velocity
of the model distribution before convolution by error.  We do not show error
bars, but the statistical errors on these points are very small. All these
points would move upwards by $20\kms$ if we increased our estimate of the
local circular speed from $\Theta_0=220\kms$ to $\Theta_0=240\kms$, and they
would move down by $5\kms$ if we decreased our estimate of
$v_{\phi\odot}-\Theta_0$ from
$12.24\kms$ to $7.24\kms$. In \figref{fig:gvbar} the points for giants show a clear trend for
$\ex{v_\phi}$ to decline with distance from the plane, as we expect given
that along this sequence $\sigma_1$  rises and increases the
asymmetric drift $v_{\rm a}\sim \sigma_1^2/v_c$.

In \figref{fig:dvbar} the point for hot dwarfs at $z\lta50\pc$ and $R<R_0$ is
$\sim25\kms$ larger than the corresponding point at $R>R_0$, so both points
are highly anomalous. However, the histograms for the associated bins (which
we do not show) indicate that the anomaly is not caused by small-number
statistics. The points for larger distances from the plane
lie close to the circular speed at $R<R_0$ and fall about $4\kms$ lower at
$R>R_0$. These differences could well reflect spiral structure. The points
for cool dwarfs show a slight fall with increasing distance from the
plane and a tendency to be up to $2\kms$ lower at $R>R_0$ than at $R<R_0$.
The fall in $\ex{v_\phi}$ between the plane and $0.5\kpc$ is consistent with
that of the giants.

\subsection{Moments of the $V_1$ and $V_3$ distributions}

The black points in Figs.~\ref{fig:hdRzFits} to \ref{fig:gRzFits} show, for
hot dwarfs, cool dwarfs, clump and non-clump giants respectively, the
distributions of the meridional-plane components $V_1$ and $V_3$ defined by
equations (\ref{eq:defsV1}). At the
bottom-centre of each panel the numbers in brackets give the mean values of
$R$ and $|z|$ for the stars in each bin, the standard deviation of the data
(sD), the value at this location of the relevant velocity dispersion from the
Gaussian model of Section \ref{sec:nobins} (sM), the mean velocity of the
stars in the bin (mV) and the rms measurement error for those stars (eV). The
agreement between the standard deviations of the data and the model
dispersion at the bin's barycentre is typically excellent. 

If the Galaxy were in an axisymmetric equilibrium and we were using the
correct value for the Sun's peculiar velocity, the mean velocities would all
vanish to within the discreteness noise, but they do not. All the three older populations show similar trends
in mean velocities: the means of $V_3$ tend to be negative at $R>R_0$ and
increase in absolute value away from the plane, while the mean values of
$V_1$ fall from positive to negative as one moves away from the plane with
the largest absolute values occurring for giants near the plane.
\cite{Siebert11} and \cite{Williams13} have analysed similar statistically
significant mean velocities in velocities of RAVE stars drawn from an earlier
spectral-analysis pipeline than that used here. We defer discussion of this
phenomenon until Section \ref{sec:errors}.

\begin{figure}
\centerline{\epsfig{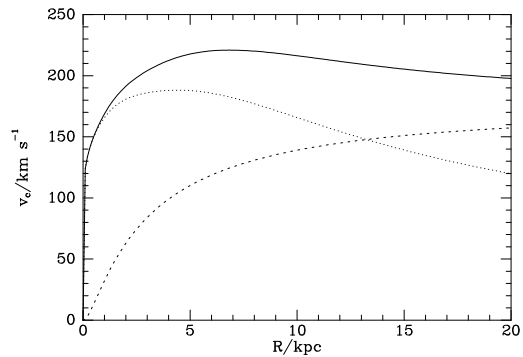}}
\caption{Dotted curve: the contribution to the circular speed from the disc
and bulge components; dashed curve: the contribution of the dark halo.}\label{fig:potential}
\end{figure}

\section{Comparisons with dynamical models}\label{sec:fjmodels}

It is interesting to compare the observed distributions with ones predicted
by the favoured equilibrium dynamical model of B12. This model is defined by
a gravitational potential and a distribution function. The potential is
generated by thin and thick exponential stellar discs, a gas layer, a
flattened bulge and a dark halo. \figref{fig:potential} shows the
contributions to the circular speed from the baryons (dotted curve) and from
the dark halo (dashed curve). One sees that this is a maximum-disc model. In
fact, 65\% of the gravitational force on the Sun is produced by baryons rather
than dark matter.

The distribution function (\df)
is an analytic function $f(\vJ)$ of the three action integrals $J_i$.
The function, which specifies the density of stars in three-dimensional
action space, has nine parameters. Four parameters specify each of the thin
and thick discs and one parameter specifies the relative weight of the thick
disc. Their values are given in column (b) of Table 2 in B12. They were
chosen by fitting the model's predictions for the velocity distribution of
solar-neighbourhood stars to that measured by the Geneva-Copenhagen survey
(GCS) of F and G stars \citep{HolmbergNA}, and to the vertical density
profile of the disc determined by \cite{GilmoreR}. Hence the data to which
this \df\ was fitted do not include velocities in the region distance
$s\gta150\pc$ within which most RAVE stars lie, and whatever success the \df\
has in {\it predicting\/} the velocities of RAVE stars must be considered a
non-trivial support for the assumptions that went into the model, which
include the use of a particular, disc-dominated, gravitational potential and
the functional form of the \df.

\begin{figure*}
\centerline{\epsfig{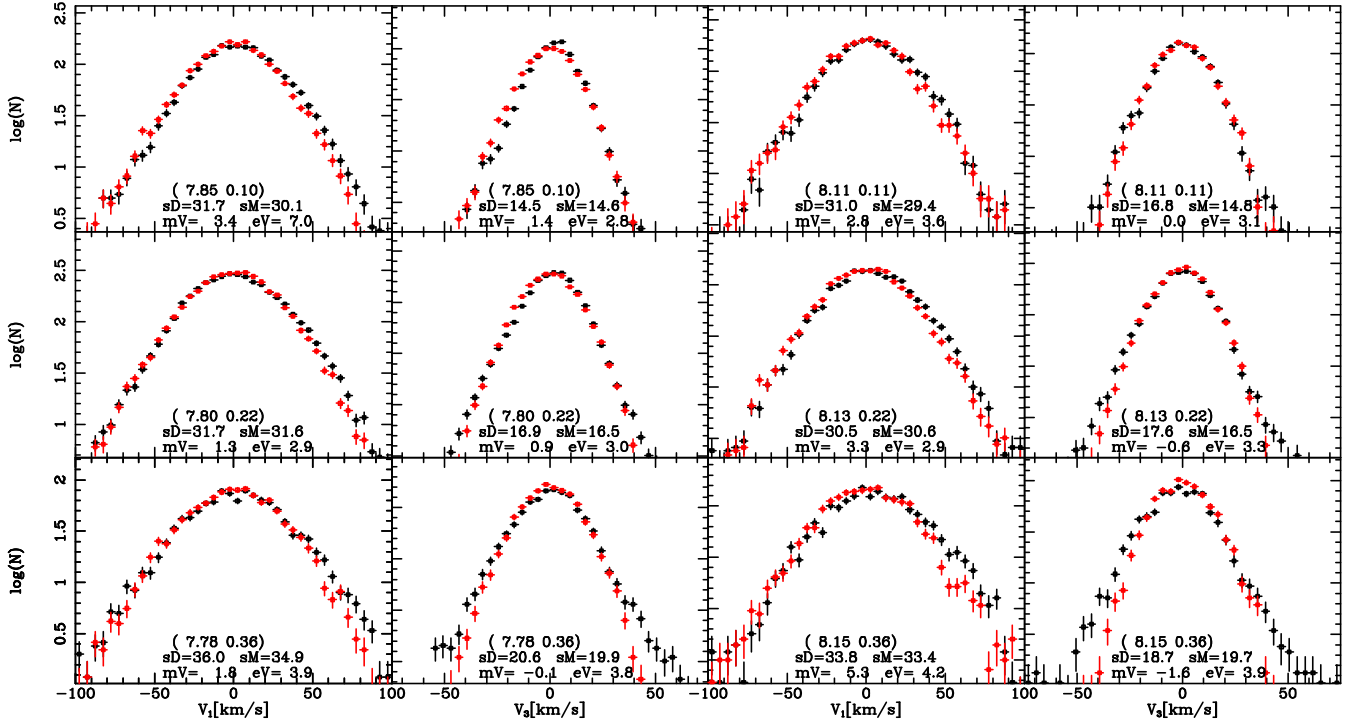}}
\caption{Distributions of $V_1\simeq -v_r$ and $V_3\simeq v_z$ for hot
dwarfs. Black points show the RAVE data, red points the predictions of the
B12 model when it is assumed that all hot dwarfs are younger than $5\Gyr$ and
as such belong to the thin disc.  At the lower middle of each panel are
given: the mean $(R,z)$ coordinates of the bin; the standard deviation of the
data after correction for error and the velocity dispersion at the mean
coordinates of the Gaussian-model described in Section \ref{sec:nobins}; the
mean of the data and the rms error of the velocities.
}\label{fig:hdRzFits}
\end{figure*}

\begin{figure*}
\centerline{\epsfig{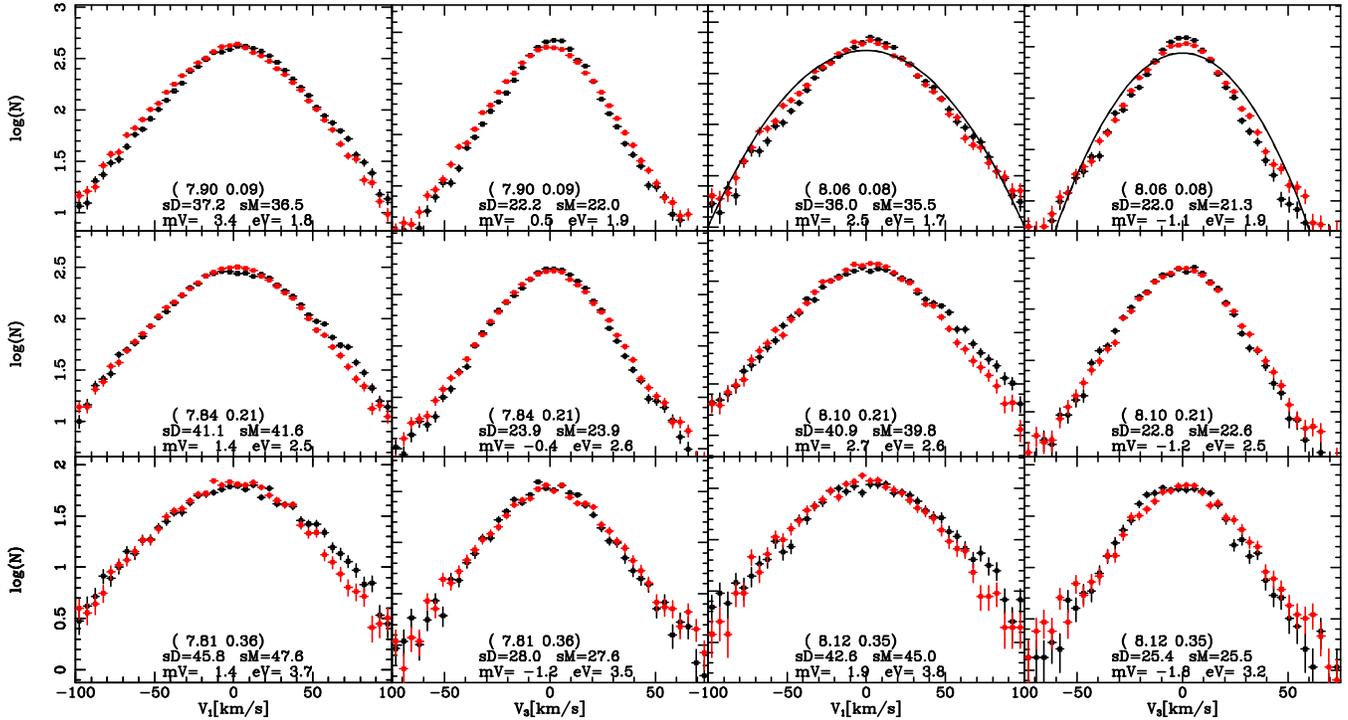}}
\caption{As Fig.~\ref{fig:hdRzFits} but for cool dwarfs. The red points now
show the predictions of the B12 model when cool dwarfs are assumed to sample
the entire \df. In the last two panels of the top row we show the Gaussian
distributions that were fitted in Section \ref{sec:nobins} to illustrate how
well the dynamical model captures the deviations of the observed distribution
from Gaussianity.
}\label{fig:cdRzFits}
\end{figure*}

We have used the B12 \df\ to generate pseudo-data for each star in the RAVE
sample from the model's velocity distribution as follows. We start by
choosing a possible true location $\vx'$ by picking a distance $s'$ from the
multi-Gaussian model of the star pdf in distance $s$ that
\cite{Binneyetal13} produced. We
then sample the velocity distribution of the dynamical model for that class
of star at $\vx'$ and compute the corresponding proper motions and
line-of-sight velocity $\vlos$. To these observables we add random errors
drawn from the star's catalogued error distributions, and from the modified
observables compute the space velocity using the catalogued distance $s$
rather than the hypothesised true distance $s'$. This procedure comes very
close to reproducing the data that would arise if the Galaxy were correctly
described by the model, each star's distance pdf were sound and the errors on
the velocities had been correctly assessed: it does not quite achieve this
goal on account of a subtle effect, which is costly to allow for. This effect
causes the procedure to over-weight slightly the possibility that stars lie
at the far ends of their distance pdfs (Sanders \& Binney in preparation). We
believe the impact of this effect to be small, so our model histograms
correctly represent the model's predictions for a survey with the selection
function and errors of RAVE.

We assume that the hot dwarfs are all younger than $5\Gyr$ (e.g., Fig.~2 of
\citealt{Zwitter10}) and
correspondingly restrict the B12 \df\ of these objects to the portion of the
thin disc that is younger than $5\Gyr$. The distributions of clump and
non-clump 
giants and cool dwarfs are (rather arbitrarily) assumed to sample the whole
\df.

\begin{figure*}
\centerline{\epsfig{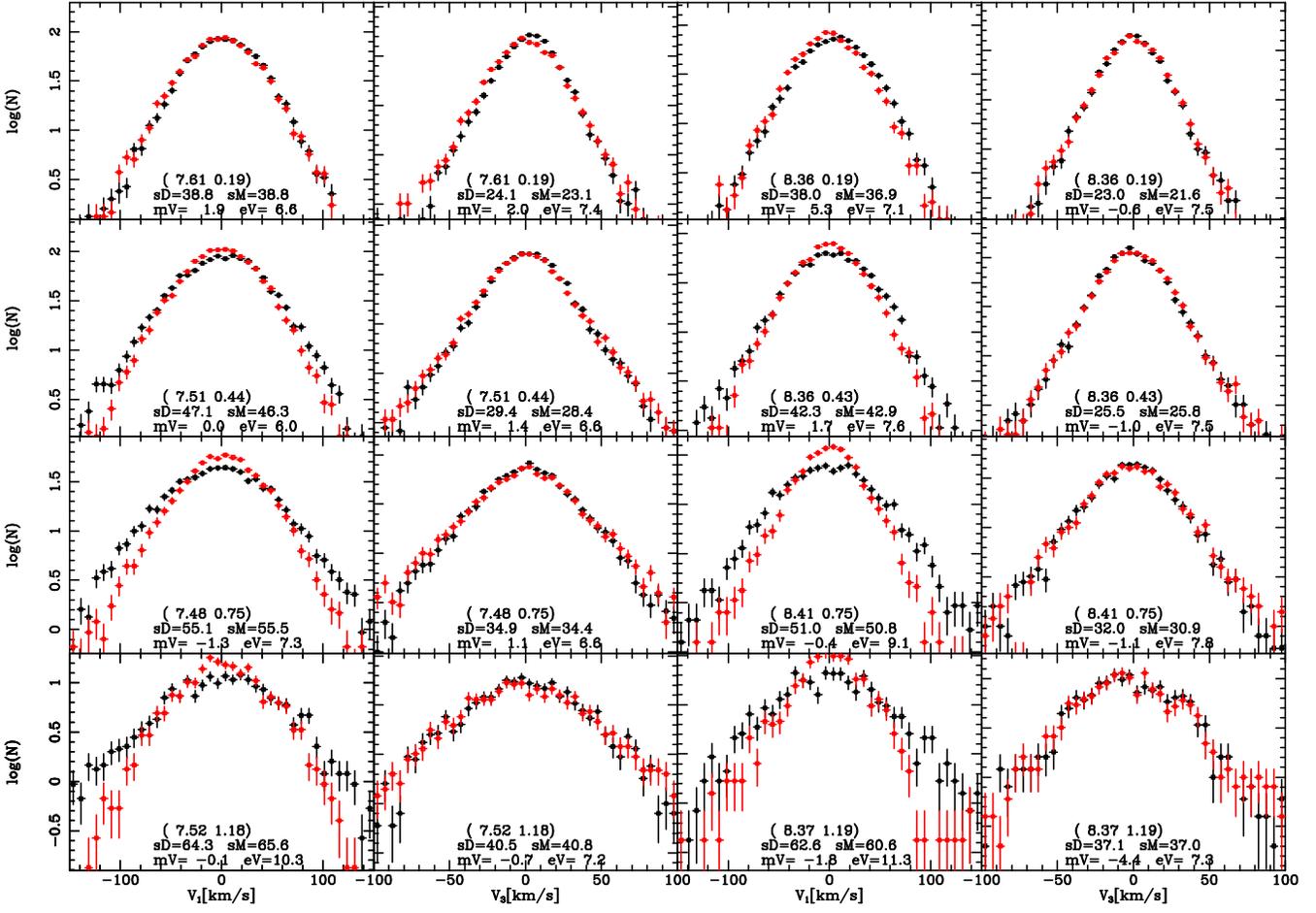}}
\caption{As \figref{fig:cdRzFits} but for clump giants.
}\label{fig:rcRzFits}
\end{figure*}

\begin{figure*}
\centerline{\epsfig{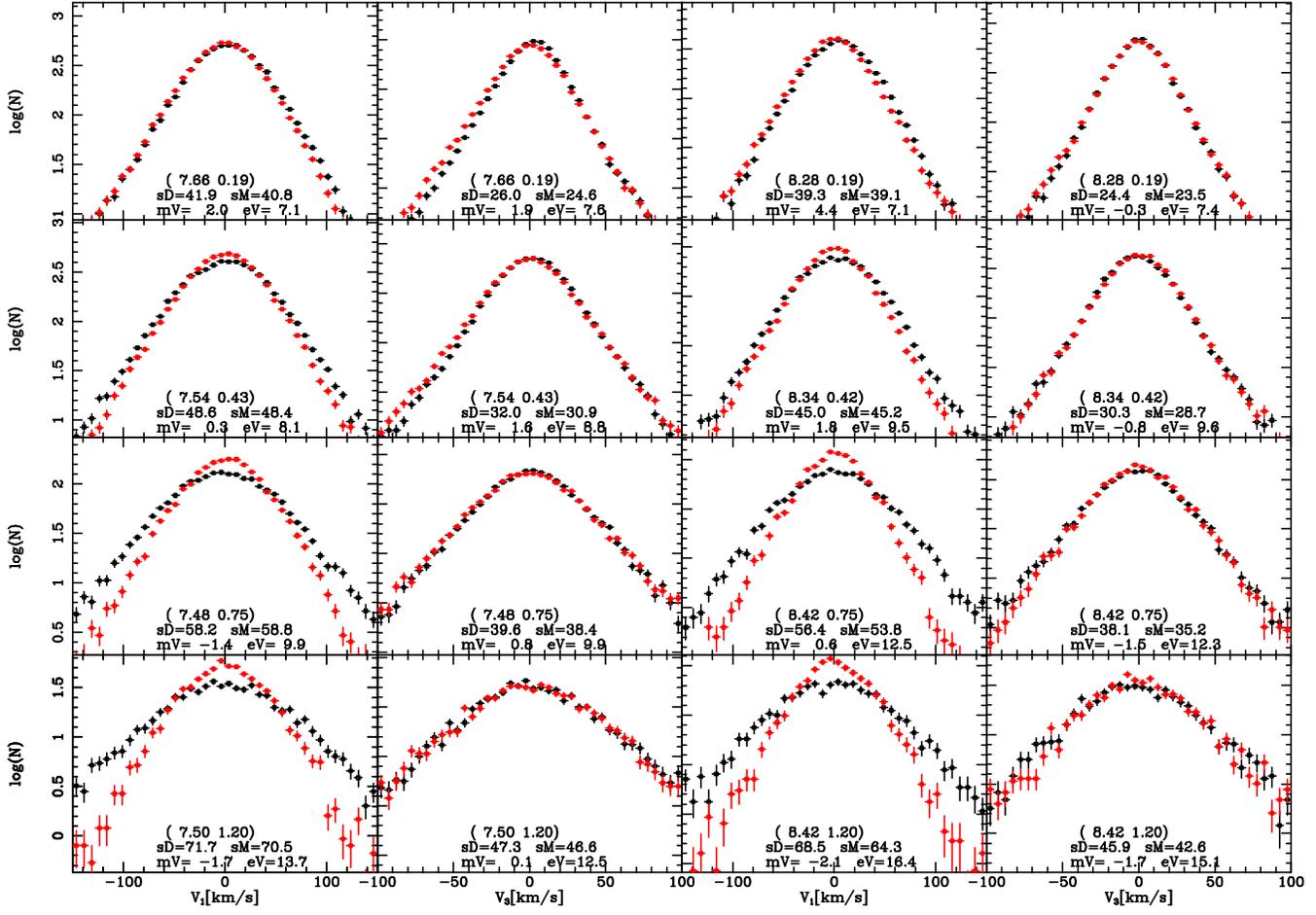}}
\caption{As \figref{fig:cdRzFits} but for non-clump giants.
}\label{fig:gRzFits}
\end{figure*}

\subsection{Azimuthal velocities distributions}

The red points in Figs~\ref{fig:rcVphiFits} to \ref{fig:cdVphiFits} show the
model's predictions for the $v_\phi$ components. Figs \ref{fig:rcVphiFits}
and \ref{fig:gVphiFits} show that the velocities of the clump giants are very
similar to those of the non-clump giants. This result is in line
with expectations, but serves to increase our confidence in our distance
estimates for, as we shall see in Section~\ref{sec:errors}, systematic errors
in the distances of whole groups of stars distort the derived velocity
distributions. Hence consistency between the histograms for clump and
non-clump giants suggests that our distances to non-clump giants, which are the
hardest to determine, are no more in error than are the distances to clump
giants. 

In Figs~\ref{fig:rcVphiFits} and \ref{fig:gVphiFits} the models definitely
under-populate the wing at $v_\phi>\Theta_0$, especially away from the plane.
This is likely to reflect the model's thick disc being radially too cool, as
discussed below.

A notable difference between the observed and predicted distributions for
both the giants and the hot dwarfs (Figs~\ref{fig:rcVphiFits} to
\ref{fig:hdVphiFits}) is that at $R<R_0$ and $|z|\sim0.5\kpc$ the black,
measured, distribution is shifted to larger values of $v_\phi$ than the red
predicted one. In the case of the hot dwarfs, a similar but distinctly
smaller shift is seen at $R>R_0$. The smaller shift at $R>R_0$ is clearly
connected to the fact that in \figref{fig:dvbar} the $\ex{v_\phi}$ points for
$R>R_0$ lie below those for $R<R_0$. At $z<0.5\kpc$ the same phenomenon is
evident for giants in \figref{fig:gvbar}. One possible explanation is that the
Galaxy's circular-speed curve is falling with $R$ relative to that of the
model.

While the theoretical distribution depends only on the model's value
$220\kms$ for the local circular speed $\Theta_0$, the observed velocities
have been derived using both $\Theta_0$ and a value $V_0=12.24\kms$ from
\cite{SchoenrichBD} for the amount by which the Sun's $v_\phi$ exceeds
$\Theta_0$. Hence an offset between the red and black curves in
Figs~\ref{fig:rcVphiFits} to \ref{fig:cdVphiFits} can be changed by changing
the assumed value of $V_0$: reducing $V_0$ shifts the black distribution to
the left.  However, the case for such a change is less than unconvincing
because the shift is clear only at $R<R_0$ and $|z|\lta0.5\kpc$. Moreover in \figref{fig:cdVphiFits} for the
cool dwarfs the model histograms provide excellent fits to the data. In
\figref{fig:hdVphiFits} for the hot dwarfs the offset between the red and
black histograms vanishes at $R>R_0$ near the plane but grows with $|z|$.

A more convincing case can be made for an increase in the width of
the theoretical distributions of giants away from the plane. 

In addition to a possibly incorrect value of $V_0$, there are four other obvious
sources of offsets between the observational and theoretical distributions of
$v_\phi$:

\begin{itemize}

\item Spiral arms must generate fluctuations in the mean azimuthal velocity
of stars. Judging by oscillations with Galactic longitude in the observed
terminal velocity of interstellar gas \cite[e.g.][]{Malhotra}, the magnitude
of this effect is probably at least as great as $7\kms$ in a population such
as hot dwarfs that has a low velocity dispersion. Moreover, it is now widely
accepted that the irregular distribution of Hipparcos stars in the $(U,V)$
plane of velocities \citep{Dehnen98} is in large part caused by spiral arms
perturbing the orbits of stars
\citep{Desimone,Antoja11,Siebert12,McMillan13}. The large (up to $20^\circ$)
value of the vertex deviation for hot dwarfs is surely also due to spiral
structure.  Spiral-induced modulations in $\ex{v_\phi}$ will vary quite
rapidly with radius and thus could make significantly different contributions
to $\ex{v_\phi}$ in our bins at $R<R_0$ and $R>R_0$.

\item The mean age of the stellar population is expected to decrease with
increasing Galactocentric distance. Such a decrease would introduce a bias
into a sample selected to be young such that there were more stars seen near
pericentre than near apocentre than in a sample of older stars, so stars in
the younger sample would tend to have larger values of $v_\phi$ than stars
in the older sample. This effect could explain why the histograms for hot
dwarfs show larger offsets than do those for cool dwarfs.

\item We are probably using a value of $R_0$ that is too small by $\sim3\%$.
Changing the adopted value of $R_0$ changes the supposed direction of the
tangential vector $\ve_\phi(\star)$ at the location of a star and thus
changes the component of a star's Galactocentric velocity $\vv$ that we deem
to be $v_\phi$. The velocity $\vv$ is made up of the star's heliocentric
velocity $\vv_{\rm h}$ and the Sun's largely tangential velocity
$\vv_\odot=\Theta_0\ve_\phi(\odot)+(U_0,V_0,W_0)$. For a star at a given
distance, increasing $R_0$ diminishes the angle between $\ve(\star)$ and
$\ve(\odot)$, and thus, by diminishing the angle between $\ve_\phi(\star)$
and $\vv_\odot$, tends to increase $v_\phi$. Consequently, in Figs
\ref{fig:rcVphiFits} to \ref{fig:cdVphiFits} increasing $R_0$ moves the black
points to the right, away from the model's predictions.

\item We are probably using a value of $\Theta_0$ that is too small by
$\sim9\%$.  Increasing $\Theta_0$ by $\delta\Theta$ simply moves the
observational histogram to the right by $\delta\Theta$.  However, since the
asymmetric drift $v_{\rm a}$ of a population that has radial velocity
dispersion $\sigma_r$ scales as $\sigma_r^2/\Theta_0$, increasing
$\Theta_0$ moves the theoretical histogram to the right by
\[
\delta\Theta-\delta v_{\rm
a}=\left(1+{\sigma_r^2\over\Theta_0^2}\right)\delta\Theta,
\]
 so this upward revision will reduce by
 $(\sigma_r/\Theta_0)^2\delta\Theta_0\sim0.04\delta\Theta_0$
the offsets we obtained with our traditional choices of $R_0$ and $\Theta_0$.

\end{itemize}

\begin{figure*}
\centerline{\epsfig{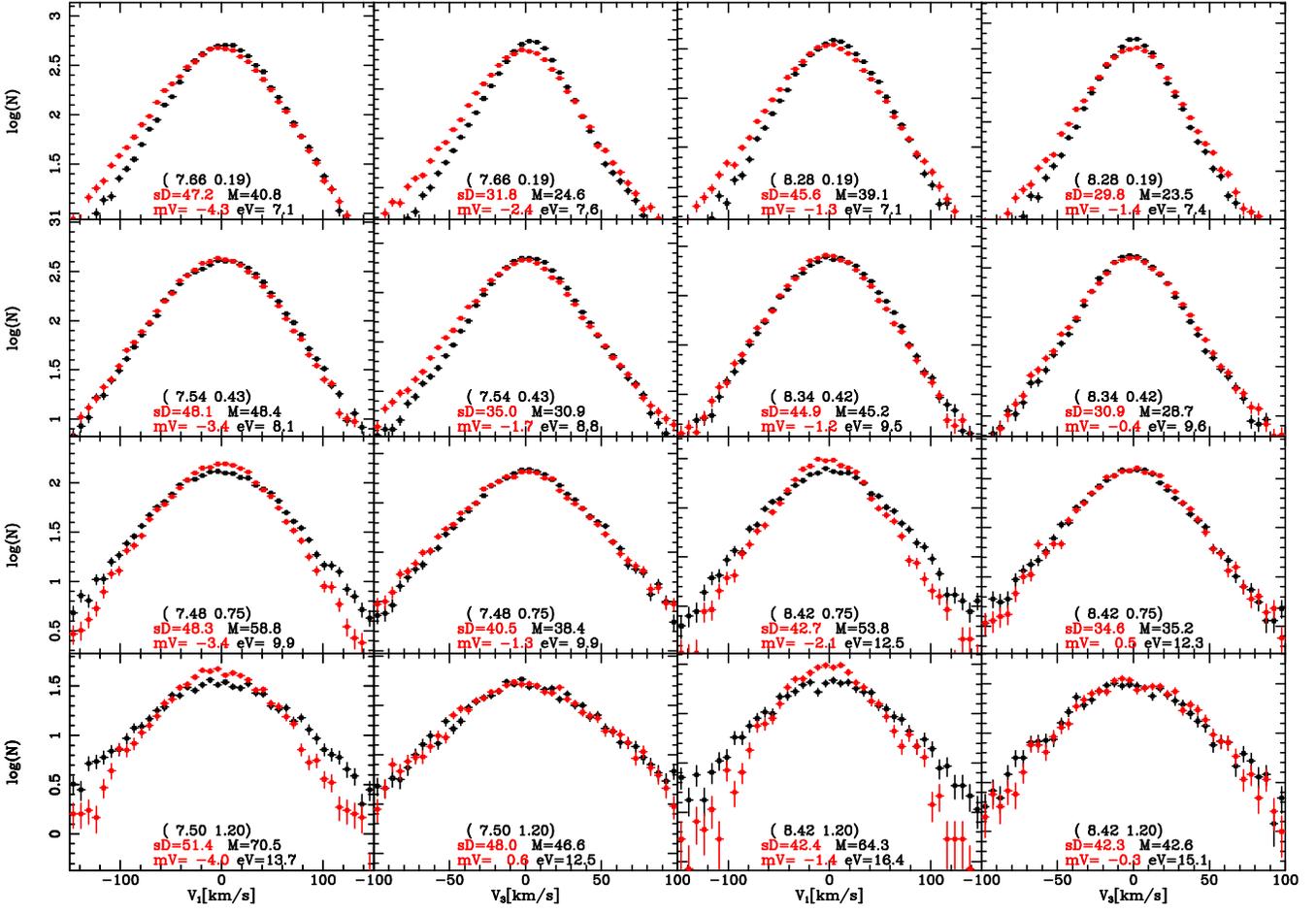}}
 \caption{The black points and curves are identical to those plotted in
\figref{fig:gRzFits}. The red model histograms have been modified by
supposing that the catalogued distance to each (giant) star is $20\%$ larger
than it should be. The values sD and mV given at the
bottom are now the standard deviation and mean of the red histogram.
}\label{fig:egRzFits}
\end{figure*}

\begin{figure}
\centerline{\epsfig{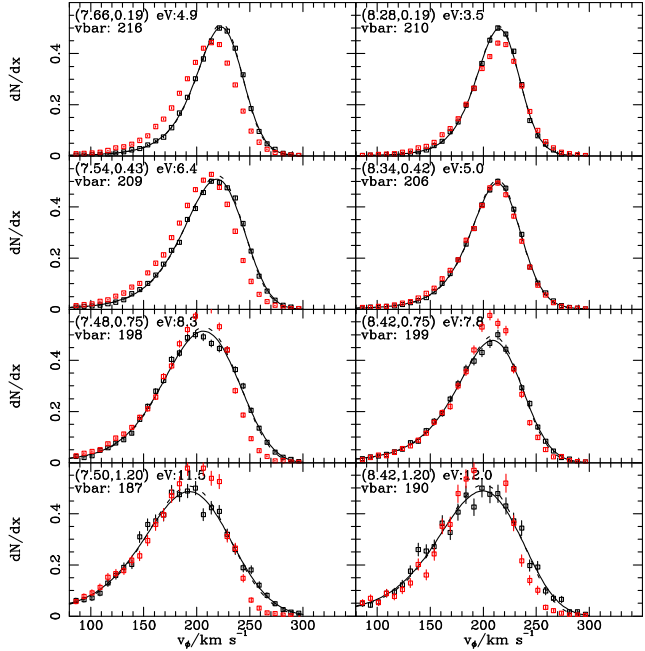}}
\caption{As \figref{fig:gVphiFits} but when the adopted distances to these (giant) stars
are $20\%$ larger than they should be.
}\label{fig:egVphiFits}
\end{figure}

\subsection{Velocities in the meridional plane}

Figs.~\ref{fig:hdRzFits} to \ref{fig:gRzFits} are the analogues of
Figs~\ref{fig:rcVphiFits} to \ref{fig:cdVphiFits} for components of velocity
$V_1$ and $V_3$ (equation \ref{eq:defsV1}) in the meridional plane: black points show observational
histograms and red ones the predictions of the B12 model. $V_1$ is the
component of velocity along the longest principal axis of the velocity
ellipsoid at the star's location according to the Gaussian model fitted in
Section \ref{sec:nobins}. The sign convention is such that at the Sun
$V_1\simeq U=-v_R$.  $V_3\simeq W=v_z$ is the perpendicular velocity
component.  The left two columns are for bins with $R<R_0$ while the right
two columns are for bins with $R>R_0$.  At the lower middle of each panel are
given: the mean $(R,z)$ coordinates of stars in the bin; the standard
deviation of the data after correction for error (sD) and the velocity
dispersion at the mean coordinates of the Gaussian-model described in Section
\ref{sec:nobins} (sM); the mean of the data (mV) and the rms error of the
velocities (eV).

All distributions are significantly non-Gaussian (i.e. the distributions are
far from parabolic) and the B12 model captures this aspect of the data
beautifully. The last two panels in the top row of \figref{fig:cdRzFits}
illustrate this phenomenon by showing the parabolas of the Gaussian
distributions fitted in Section \ref{sec:nobins}. Notwithstanding the
non-Gaussian nature of the velocity distributions, in every bin there is good
agreement between the standard deviation of the data sD and the dispersion at
of the Gaussian model sM at the barycentre of the bin. This result implies
that equations (\ref{eq:defsmod}) can be safely used to recover the principal
velocity dispersions throughout the studied region.

The model is particularly successful in predicting the $V_3$ distributions of
both dwarfs and giants.
In the case of the dwarfs, the only blemish on its $V_3$ distributions is a
marginal tendency for the distribution of hot dwarfs to be too narrow at high
$|z|$.

The principal differences between the model and observed $V_1$ distributions
of dwarfs arise from left-right asymmetries in the data. For example, in the
third panels from the left in the first and second rows of
\figref{fig:hdRzFits} for hot dwarfs, the black points lie systematically
above the red points for $V_1>0$ (inward motion), a phenomenon also evident
in the top left panel of that figure. In the first and third panels in the
second row of \figref{fig:cdRzFits} for cool dwarfs, a similar phenomenon is
evident in that the red points lie above the black points at $V_1<0$. A
contribution to these divergences must come from star streams, which
\cite{Dehnen98} showed to be prominent in the local $UV$ plane.

Figs \ref{fig:rcRzFits} and \ref{fig:gRzFits} for clump and non-clump giants
show $V_1$ and $V_3$ distributions in bins that extend to much further from
the plane. In both cases the model and observed $V_3$ distributions agree to
within the errors. Given the smallness of the error bars in the case of the
giants and the fact that the data extend to a distance from the plane that is
more than ten times the extent of the GCS data to which the B12 model was fitted,
the agreement between the observed and theoretical $V_3$ histograms in
\figref{fig:gRzFits} amounts to a very strong endorsement of the B12 model.

The observed $V_1$ distributions for clump and non-clump giants are
consistent with one another, and the superior statistics of non-clump giants
highlight the deviations from the model predictions. Near the plane the model
fits the data well, but the further one moves from the plane, the more clear
it becomes that the model distribution of $V_1$ is too narrow. This
phenomenon arises because in B12, contrary to expectation, the thick
disc needed to be radially cooler than the thin disc. The RAVE data are
indicating that this was a mistake. In B12 two factors shared responsibility
for the radial coolness of the thick disc. One was the ability of the
thin-disc \df\ to fit the wings of the $U$ and $V$ distributions in the GCS,
leaving little room for the thick disc's contribution there.
The other factor was an indication from SDSS that $\ex{v_\phi}$ does not fall
rapidly with distance from the plane. \figref{fig:gVphiFits} relates to this
second point, and indeed the RAVE data show more stars with large $v_\phi$
than the model, especially at large $|z|$. In B12 it was demonstrated that
there is a clean dynamical trade-off between  $\ex{v_\phi}$ and $\sigma_\phi$ in the
sense that an increase in the former has to be compensated by a decrease in
the latter. Moreover,  $\sigma_\phi$ is dynamically coupled to
$\ex{V_1^2}^{1/2}$, so if one is reduced the other must be reduced as well.
Hence large $\ex{v_\phi}$ implies small $\ex{V_1^2}^{1/2}$. There is a puzzle
here that requires further work.

\subsection{Effect of distance errors}\label{sec:errors}

Our model predictions already include the effects of random distance (and
velocity) errors. Now we investigate how {\it systematic\/} errors in our
spectrophotometric distances affect the derived kinematics. This
investigation is motivated in part
by the indication in \cite{Binneyetal13} from the kinematic test of
\cite{SBA} that distances to giants might be over-estimated by as much as
20\%, and distances to the hottest dwarfs under-estimated by a similar
amount. 

The black points in \figref{fig:egRzFits} are identical to those in the
corresponding panels of \figref{fig:gRzFits} but the red model points have
been modified by adding $-5\log_{10}(\e)\times0.2$ to the randomly chosen
distance modulus of each star before evaluating the \df. This modification
enables us to model the impact on the survey of catalogued distances being on
average $20$ per cent too large.  

The figure shows that such distance errors introduce left-right asymmetry
into the model distributions of both $V_1$ and $V_3$ similar to that
evident in the $V_1$ distribution of hot dwarfs. The red values of $mV$ at
the bottom middle of each panel, show the mean values of $V_1$ and $V_3$ for
the model histograms. We see that these values are non-zero and of comparable
magnitude to the mean values of the observed histograms given in
\figref{fig:gRzFits}. Thus non-zero mean values of $\ex{V_1}$ and $\ex{V_3}$
may arise from distance errors rather than from real streaming motion.
However, near
the plane our distance errors induce negative mean values of $V_1$ (net
outward motion) whereas the data histogram shows a smaller positive mean
value of $V_1$.

Physically, over-estimating distances makes the $V_1$ distribution skew to
positive $V_1$ because the survey volume is not symmetric in Galactic
longitude, and at certain Galactic longitudes proper motion generated by the
disc's differential rotational is wrongly interpreted to be proper motion
associated with motion towards the Galactic centre.

The assumption that distances are over-estimated also broadens the model
distribution of $V_1$ far from the plane, with the result that, for example,
in the third row of \figref{fig:egRzFits} the red and black points for $V_1$
lie significantly closer than in the corresponding panels of
\figref{fig:gRzFits}.  

\figref{fig:egVphiFits} is the analogue of \figref{fig:gVphiFits} for the
case in which the distances to giants have been over-estimated by $20\%$.  In
the top left panel for small $|z|$ and $R<R_0$ the agreement between model
and data is now less good than it is in \figref{fig:gVphiFits}, but in every
other panel the agreement is at least as good in \figref{fig:gVphiFits} and
for $R>R_0$ it is distinctly improved. Thus the $v_\phi$ distributions by no
means speak against the suggestion that many distances have been
over-estimated by $\sim20\%$.

While in \figref{fig:egRzFits} distance errors have improved the fit to the
data only at $|z|>0.5\kpc$ and weakened the fit closer to the plane, it is
perfectly possible that systematic errors are largely confined to more
distant stars and/or ones further from the plane. In fact, such an effect is
inevitable even if the errors in distances of individual stars were
inherently unbiased because stars that happen to pick up a positive distance
error will tend to accumulate in the distant bins, and conversely for stars
that happen to pick up a negative distance error. When we modified the
model's predictions to allow for random distance errors, we did not capture
this effect because the spatial bin to which a star is then assigned is not
affected by whether it is supposed to have had its distance over- or
under-estimated.

\section{Discussion}

\cite{Siebert11} reported a significant radial gradient in the mean
$\ex{v_R}$ of velocities of stars reduced by the RAVE VDR2 pipeline.
Williams et al.\ (2013; hereafter W13) used data from the VDR3 pipeline to
analyse the mean velocity field $\ex{\vv}$ of clump stars. In a steady-state,
axisymmetric Galaxy the only non-vanishing component of this field would be
$\ex{v_\phi}$ and it would have a maximum in the plane, falling away with
$|z|$ symmetrically on each side. Instead Fig.~11 of W13 indicates that the
velocity field of the clump stars has both $\ex{v_R}$ and $\ex{v_z}$
components non-zero and with gradients in both the $R$ and $z$ directions,
and there is a lack of symmetry about the plane. W13 strike a cautionary note
by showing that the $\ex{v_R}$ and $\ex{v_z}$ components are sensitive to
which proper motions one adopts, but they demonstrate that $\ex{\vv}$ is
insensitive to the adopted absolute magnitude of clump stars.

As W13 show, probing the observed velocity field is made difficult by the
complexity of the three-dimensional volume surveyed by RAVE: samples
assembled to have a progression of values of one coordinate inevitably differ
systematically in another coordinate as well. For this reason it is crucial
to compare observational results with the predictions of a model that suffers
the same selection effects. W13 compare the observations to mock catalogues
selected by the code {\sc Galaxia} \citep{Sharmaetal} from the Besan\c con
model \citep{Robin03}. Our comparisons differ in that (i) we have used a
fully dynamical model, based on Jeans' theorem, rather than the essentially
kinematic Besan\c con model, and (ii) we assign new velocities to existing
stars rather than drawing an entirely new sample from the model -- this
procedure has the great advantage that we do not have to engage with the
survey's complex photometric selection function. 

Our emphasis has been different in that we have focused on entire velocity
distributions rather than just the distributions' means. This has been
possible because we have a more prescriptive dynamical model, but it has
resulted in our using much bigger bins than W13. In particular, we have
grouped together stars above and below the plane, which will inevitably wash
out some of the structure in the $(R,z)$ plane seen by W13.

Our demonstration that introducing plausible systematic errors in the assumed
distances to stars causes the model histograms to acquire mean velocities
that are similar in magnitude to those found by Williams et al.\ (2013) must be a concern
even though the particular systematic in distance error that we have
considered does not generate the observed pattern of mean velocities. The
extent to which distance errors broaden the distributions of $V_1$ is
surprising and interesting given the difficulties one encounters finding a
dynamical model that is consistent with all the data for $\ex{v_\phi}$ and
$\ex{V_1^2}^{1/2}$ in the absence of systematic distance errors.

\section{Conclusions}

We have analysed the kinematics of $\sim400\,000$ RAVE stars for which
\cite{Binneyetal13} have deduced pdfs in distance modulus. The sample divides
naturally into clump and non-clump giants, hot and cool dwarfs.  For each of
these classes, and without binning the data, we have obtained analytic
formulae for the structure of the velocity ellipsoid at each point in the
$(R,z)$ plane. We are able to map the velocity ellipsoid of the giants to
distances $\sim2\kpc$ from the Sun and find that at $(R,z)$ the direction of
the longest axis is inclined to the Galactic plane by an angle
$\sim0.8\arctan(z/R)$. The lengths of the $(R,z)$ semi-axes are in the
ratio $\sigma_3/\sigma_1\simeq 0.6$. The velocity dispersions rise with
distance from the plane, from $\sigma_r\simeq37\kms$, $\sigma_z\simeq21\kms$
at $(R_0,0)$ to $\sigma_r\simeq82\kms$, $\sigma_z\simeq54\kms$ at
$(R_0,2\kpc)$.  The velocity ellipsoid of the cool dwarfs cannot be traced to
great distances, but it is consistent with being the same as that of the
giants. In the plane the velocity dispersions of the hot dwarfs are
$\sigma_r\simeq29\kms$ and $\sigma_z\simeq14\kms$ and they increase rather
slowly with distance from the plane. From equations (\ref{eq:defa0}) and
(\ref{eq:defsmod}) and Table~\ref{tab:SigzSigR} one can compute for any of
our four classes of star the structure of the velocity ellipsoid at a general
point in the $(R,z)$ plane.

We have used a novel formula to obtain remarkably precise analytic fits to
the distinctly non-Gaussian $v_\phi$ distributions for eight bins in the
$(R,z)$ plane. The complete $v_\phi$ distributions at these points can be
recovered for any of the four classes of stars by inserting values from
either Table \ref{tab:gVphi} or Table \ref{tab:dVphi} into equations
equations (\ref{eq:Pphi}) and (\ref{eq:Svphi}).

We have compared our observational velocity histograms with the predictions of
a dynamical model that was fitted to the local velocity distribution and the
\cite{GilmoreR} vertical density profile. When making this comparison we
assume only that the survey's selection function is velocity-blind (which it
certainly is) and we are able to model the effects of errors in both
distances and velocities with considerable completeness.

Overall the agreement between the model's predictions and the data is
remarkably good and offers strong support for the assumptions on which the
dynamical model rests, including its gravitational potential. There is,
however, a tendency for the distribution of observed $v_\phi$ components to
be shifted to larger values than the model predicts. A possible contributory
factor to this offset may be over-estimation of the Sun's peculiar $V$
velocity, but the offset can be generated in several ways, including spiral
arms, the age gradient within the disc, and use of incorrect values of $R_0$
and $\Theta_0$.

The dynamical model performs outstandingly well in predicting the
distributions of vertical velocity components $V_3$ of all star classes.
These distributions are considerably more sharply peaked than Gaussians and
the model captures this phenomenon beautifully. At $|z|<0.5\kpc$ the model
predicts the distributions of radial components $V_1$ nearly as successfully,
but at greater distances from the plane the model predicts distributions of
$V_1$ that are too narrow. This problem is undoubtedly connected to the
surprising conclusion of B12 that the thick disc is radially cooler than the
thin disc, a conclusion driven by both the structure of the GCS histograms
for $U$ and the strong mean rotation of SDSS stars far from
the plane. The RAVE data also require that at $|z|>1\kpc$ there are
unexpectedly many stars at large $v_\phi$, and this fact constraints our
ability to make the thick disc radially hotter as the $V_1$ histograms imply.

One way to resolve, or at least ameliorate, the problem is to suppose that
stars in the most distant bins have had their distances over-estimated by
$\sim20\%$.  Similar distance over-estimates in the nearer bins would impair
the nice agreement between theory and observation. However, it is inevitable
that stars placed in the most distant bins have, on average, over-estimated
distances, so it is plausible that distance over-estimates contribute
significantly to the anomalies in the high-$|z|$ bins.

This study clearly indicates that the approach to Galaxy modelling developed
in B12 is well worth developing. There are several
directions in which to go. First a new \df\ of the current type should be
fitted to the richer body of observational data that is now available using
an updated Galactic potential $\Phi$. Next this \df\ and these data should be
used as a starting point for a re-determination of $\Phi$ along the lines
outlined by \cite{McMillanB13}. Currently the \df\ is being extended
to include chemistry alongside age \citep{BinneyS13}: this extension should
markedly increase our ability to diagnose $\Phi$ because the requirement that
several stellar populations that differ in both their chemistry and their
kinematics exist harmoniously in a common potential will strongly constrain
$\Phi$.

\section*{Acknowledgements}

We thank P.J. McMillan for valuable comments on the manuscript.

Funding for RAVE has been provided by: the Australian Astronomical Observatory;
the Leibniz-Institut f\"ur Astrophysik Potsdam (AIP); the Australian
National University; the Australian Research Council; the French National Research
Agency; the German Research Foundation (SPP 1177 and SFB 881); the
European Research Council (ERC-StG 240271 Galactica); the Istituto Nazionale
di Astrofisica at Padova; The Johns Hopkins University; the National Science
Foundation of the USA (AST-0908326); the W. M. Keck foundation; the Macquarie
University; the Netherlands Research School for Astronomy; the Natural
Sciences and Engineering Research Council of Canada; the Slovenian Research
Agency; the Swiss National Science Foundation; the Science \& Technology
Facilities Council of the UK; Opticon; Strasbourg Observatory; and the
Universities of Groningen, Heidelberg and Sydney. The RAVE web site is at
http://www.rave-survey.org.

\label{lastpage}
\end{document}